\documentclass[11pt,a4paper]{article}

\usepackage{epsfig,amsmath,amssymb,cite}

\begin{document}

\title{Spherical thin shells in $F(R)$ gravity: construction and stability} 
\author{Ernesto F. Eiroa$^{1, 2}$\thanks{e-mail: eiroa@iafe.uba.ar}, Griselda Figueroa Aguirre$^{1}$\thanks{e-mail: gfigueroa@iafe.uba.ar}\\
{\small $^1$ Instituto de Astronom\'{\i}a y F\'{\i}sica del Espacio (IAFE, CONICET-UBA),}\\
{\small Casilla de Correo 67, Sucursal 28, 1428, Buenos Aires, Argentina}\\
{\small $^2$ Departamento de F\'{\i}sica, Facultad de Ciencias Exactas y 
Naturales,} \\ 
{\small Universidad de Buenos Aires, Ciudad Universitaria Pabell\'on I, 1428, 
Buenos Aires, Argentina}} 

\maketitle

\begin{abstract}

We present a broad class of spherical thin shells of matter in $F(R)$ gravity. We show that the corresponding junction conditions determine the equation of state between the energy density and the pressure/tension at the surface. We analyze the stability of the static configurations under perturbations preserving the symmetry. We apply the formalism to the construction of charged bubbles and we find that there exist stable static configurations for a suitable set of the parameters of the model.\\


\end{abstract}

\section{Introduction}\label{intro} 

The Darmois--Israel \cite{daris} formalism provides the tools to analyze the characteristics and dynamics of thin shells of matter in General Relativity; it relates the energy--momentum tensor of a shell with the geometries of the regions at both sides of it. This formalism has been broadly applied in many different contexts because of its flexibility and simplicity; the stability analysis is easy to perform in highly symmetric situations, and the matter can be confined to the shell. Many researchers have adopted this formalism to model vacuum bubbles and thin layers around black holes \cite{sh1,sh2,sh3}, wormholes \cite{wh1,wh2,whcil}, and gravastars \cite{gravstar}, among others.  

The accelerated expansion of the Universe, the rotation curves of galaxies, and the anisotropy of the microwave background radiation can be explained within the context of General Relativity by adopting non--standard fluids, such as dark matter and dark energy. However, in order to avoid the use of these fluids, other approaches can be adopted, such as modified gravity; one of them corresponds to the so-called $F(R)$ theories \cite{dft,nojod} in which the Einstein-Hilbert Lagrangian is replaced by a function $F(R)$ of the Ricci scalar $R$. The junction conditions in this theory \cite{dss,js1} are more stringent than in General Relativity. For non--linear $F(R)$, at the matching hypersurface the continuity of the trace of the second fundamental form is always required and the continuity of the curvature scalar $R$ is also required, except in the quadratic case \cite{js1}. In quadratic $F(R)$ gravity, the surface has, in general, in addition to the standard energy--momentum tensor, an external energy flux vector, an external scalar pressure (or tension), and another energy--momentum contribution resembling classical dipole distributions. In order to have a divergence--free energy--momentum tensor, which guarantees local conservation, all these contributions have to be present \cite{js1,js2-3}. These results were recently extended to any quadratic theory lagrangian \cite{js4}. Within $F(R)$ gravity, several studies have been performed in recent years such as static and spherically symmetric black holes \cite{bhfr1,bhfr2,bhfr3,dft}, traversable wormholes \cite{whfr,eirfig} and pure double layer bubbles \cite{bb1}.

In this article, we construct a family of spherical thin shells by using the junction formalism in $F(R)$ gravity theories and we analyze the stability of the static configurations under radial perturbations. In order to provide concrete examples, we consider bubbles which are characterized by having an inner vacuum region separated by a thin layer of matter from an outer region.
In Sec. 2, we study geometries with constant curvature scalar $R_0$ at both sides of the shell. In Sec. 3, we analyze the quadratic case with  $R_1\neq R_2$, both constants and corresponding to the inner and the outer parts of the spacetime, respectively. In Sec. 4, we apply the equations obtained to the construction of charged bubbles. Finally, in Sec. 5, we show the conclusions of this work. We use units so that $G=c=1$, with $G$ the gravitational constant and $c$ the speed of light.

\section{Spacetimes with constant curvature scalar}\label{tswh}

We start by considering a manifold composed of two regions with the same constant curvature scalar, separated by a thin shell of matter.

\subsection{Thin shell construction}

We take two different spherically symmetric solutions in $F(R)$ gravity, with metrics
\begin{equation} 
ds^2=-A_{1,2} (r) dt^2+A_{1,2} (r)^{-1} dr^2+r^2(d\theta^2 + \sin^2\theta d\phi^2),
\label{metric}
\end{equation}
where $r>0$ is the radial coordinate, and $0\le \theta \le \pi$ and $0\le \varphi<2\pi $ are the angular coordinates. By using the junction formalism in $F(R)$ gravity, we proceed with the construction of a new manifold by selecting a radius $a$ and cutting two regions $\mathcal{M}_1$ and $\mathcal{M}_2$ defined as the inner $0\le r \leq a$ and the outer $r\geq a$ parts of the geometries 1 and 2, respectively. These regions are pasted to one another at the surface $\Sigma $ with radius $a$. This construction results in the spacetime $\mathcal{M}=\mathcal{M}_1 \cup \mathcal{M}_2$, with the inner zone corresponding to $\mathcal{M}_1$ and the exterior one to $\mathcal{M}_2$. The jump across  $\Sigma$ of any quantity $\Upsilon $ is defined as $[\Upsilon ]\equiv (\Upsilon ^{2}-\Upsilon  ^{1})|_\Sigma $. We denote the unit normals at the surface $\Sigma$ by $n^{1,2}_\gamma$ (pointing from $\mathcal{M}_1$ to $\mathcal{M}_2$), the first fundamental form by $h_{\mu \nu}$, and the second fundamental form (or extrinsic curvature) by $K_{\mu \nu}$.

Let us review the junction formalism in $F(R)$ gravity theories. In this case, there exist several conditions that should be fulfilled by our construction. One of them is the continuity of the first fundamental form i.e.  $[h_{\mu \nu}]=0$, ensuring in this way that $\mathcal{M}$ is geodesically complete. Another one is the continuity of the trace of the second fundamental form i.e. $[K^{\mu}_{\;\; \mu}]=0$. When $F'''(R) \neq 0$ (the prime on $F(R)$ means the derivative with respect to the curvature scalar $R$) the continuity of $R$ across the surface $\Sigma$ is also required i.e.  $[R]=0$.
The field equations at $\Sigma$ in this case take the form \cite{js1}
\begin{equation} 
\kappa S_{\mu \nu}=-F'(R)[K_{\mu \nu}]+ F''(R)[\eta^\gamma \nabla_\gamma R]  h_{\mu \nu}, \;\;\;\; n^{\mu}S_{\mu\nu}=0,
\label{LanczosGen}
\end{equation}
where $\kappa =8\pi $ and $S_{\mu \nu}$ represents the energy--momentum tensor at the shell. If $F'''(R) = 0$, the curvature scalar can be discontinuous at $\Sigma $, and the field equations read \cite{js1}
\begin{equation}
\kappa S_{\mu \nu} =-[K_{\mu\nu}]+2\alpha \left( [n^{\gamma }\nabla_{\gamma}R] h_{\mu\nu}-[RK_{\mu\nu}] \right), \;\;\;\; n^{\mu}S_{\mu\nu}=0;
\label{LanczosQuad}
\end{equation}
there are also three other contributions:
an external energy flux vector
\begin{equation}
\kappa\mathcal{T}_\mu=-2\alpha \bar{\nabla}_\mu[R],  \qquad  n^{\mu}\mathcal{T}_\mu=0,
\label{Tmu}
\end{equation}
where $\bar{\nabla }$ is the intrinsic covariant derivative on $\Sigma$, an external scalar pressure or tension
\begin{equation}
\kappa\mathcal{T}=2\alpha [R] K^\gamma{}_\gamma ,
\label{Tg}
\end{equation}
and a two-covariant symmetric tensor distribution 
\begin{equation}
\kappa \mathcal{T}_{\mu \nu}=\nabla_{\gamma } \left( 2\alpha [R] h_{\mu \nu } n^{\gamma } \delta ^{\Sigma }\right),
\label{dlay1}
\end{equation}
with $\delta ^{\Sigma }$ the Dirac delta on $\Sigma $, or equivalently 
\begin{equation}
\kappa \left<\mathcal{T}_{\mu \nu},\Psi ^{\mu \nu } \right> = -\int_\Sigma 2\alpha[R] h_{\mu \nu }  n^\gamma\nabla_\gamma \Psi ^{\mu \nu },
\label{dlay2}
\end{equation}
for any test tensor field $\Psi ^{\mu\nu}$. 
In quadratic $F(R)$, besides the standard energy--momentum tensor $S_{\mu \nu}$, the shell can have an external energy flux vector $\mathcal{T} _{\mu}$, an external scalar pressure/tension $\mathcal{T} $, and a double layer energy--momentum contribution $\mathcal{T}_{\mu \nu}$ of Dirac ``delta prime'' type, resembling classical dipole distributions. All these contributions are necessary in order to ensure the energy--momentum tensor to be divergence--free, a condition that allows for conservation locally  \cite{js1}. In non--linear $F(R)$ theory, the conditions for proper matching without a thin shell are more demanding than in General Relativity, besides $[h_{\mu \nu}]=0$ and $[K_{\mu \nu}]=0$, the relations $[R]=0$ and $[\nabla_{\gamma}R]=0$ are also required \cite{js1}.

We first analyze the scenario with constant curvature scalar $R_0$ at both sides of $\Sigma$, so that the condition $[R]=0$ is automatically fulfilled, and Eqs. (\ref{LanczosGen}) and (\ref{LanczosQuad}) both take the form
\begin{equation} 
\kappa S_{\mu \nu}=-F'(R_0)[K_{\mu \nu}];
\label{fieldeq}
\end{equation}
in the quadratic case  $\mathcal{T} $, $\mathcal{T} _{\mu}$ and $\mathcal{T}_{\mu \nu}$ are all zero because they are proportional to $[R]$. 
On the surface $\Sigma $ we adopt the coordinates $\xi ^{i}=(\tau ,\theta,\varphi )$, with $\tau $ the proper time on the shell. The radius $a(\tau)$ is, from now on, a function of the proper time. 
The first fundamental form associated with the two sides of the shell is
\begin{equation}
h^{1,2}_{ij}= \left. g^{1,2}_{\mu\nu}\frac{\partial X^{\mu}_{1,2}}{\partial\xi^{i}}\frac{\partial X^{\nu}_{1,2}}{\partial\xi^{j}}\right| _{\Sigma },
\end{equation}
and the second fundamental form  is given by
\begin{equation}
K_{ij}^{1,2 }=-n_{\gamma }^{1,2 }\left. \left( \frac{\partial ^{2}X^{\gamma
}_{1,2} } {\partial \xi ^{i}\partial \xi ^{j}}+\Gamma _{\alpha \beta }^{\gamma }
\frac{ \partial X^{\alpha }_{1,2}}{\partial \xi ^{i}}\frac{\partial X^{\beta }_{1,2}}{
\partial \xi ^{j}}\right) \right| _{\Sigma },
\label{sff}
\end{equation}
where the unit normals ($n^{\gamma }n_{\gamma }=1$) are
\begin{equation}
n_{\gamma }^{1,2 }=\left\{ \left. \left| g^{\alpha \beta }_{1,2}\frac{\partial G}{\partial
X^{\alpha }_{1,2}}\frac{\partial G}{\partial X^{\beta }_{1,2}}\right| ^{-1/2}
\frac{\partial G}{\partial X^{\gamma }_{1,2}} \right\} \right| _{\Sigma },
\end{equation}
in which the function $G(r)=r-a$ is zero at $\Sigma$. We adopt the orthonormal basis $\{ e_{\hat{\tau}}=e_{\tau }, e_{\hat{\theta}}=a^{-1}e_{\theta }, e_{\hat{\varphi}}=(a\sin \theta )^{-1} e_{\varphi }\} $ at the shell for the geometry (\ref{metric}). 
Within this frame, the first fundamental form is $h^{1,2}_{\hat{\imath}\hat{\jmath}}= \mathrm{diag}(-1,1,1)$, the unit normals are
\begin{equation} 
n_{\gamma }^{1,2}= \left(-\dot{a},\frac{\sqrt{A_{1,2}(a)+\dot{a}^2}}{A_{1,2}(a)},0,0 \right),
\end{equation}
and the second fundamental form is given by
\begin{equation} 
K_{\hat{\theta}\hat{\theta}}^{1,2}=K_{\hat{\varphi}\hat{\varphi}}^{1,2
}=\frac{1}{a}\sqrt{A_{1,2} (a) +\dot{a}^2}
\label{e4}
\end{equation}
and
\begin{equation} 
K_{\hat{\tau}\hat{\tau}}^{1,2 }=-\frac{A '_{1,2}(a)+2\ddot{a}}{2\sqrt{A_{1,2}(a)+\dot{a}^2}},
\label{e5}
\end{equation}
with the prime on $A_{1,2}(r)$ representing the derivative with respect to $r$. By using Eqs. (\ref{e4}) and (\ref{e5}), the condition $[K^{\hat{\imath}}_{\;\; \hat{\imath}}]=0$ reads
\begin{equation} 
-\frac{2a\ddot{a}+a A_{1}'(a) + 4(A_{1}(a)+\dot{a}^2)}{\sqrt{A_{1}(a)+\dot{a}^2}}+\frac{2a\ddot{a}+a A_{2}'(a) + 4(A_{2}(a)+\dot{a}^2)}{\sqrt{A_{2}(a)+\dot{a}^2}}=0.
\label{CondGen}
\end{equation}
Considering that the stress-energy tensor in the orthonormal basis has the form $S_{_{\hat{\imath}\hat{\jmath} }}={\rm diag}(\sigma ,p_{\hat{\theta}},p_{\hat{\varphi}})$ with $\sigma$ the surface energy density and $p_{\hat{\theta}}=p_{\hat{\varphi}}=p$ the transverse pressures, we obtain
\begin{equation} 
\sigma= \frac{F'(R_0)}{2\kappa}\left( \frac{2\ddot{a}+A_{2}'(a)}{\sqrt{A_{2}(a)+\dot{a}^2}}-\frac{2\ddot{a}+A_{1}'(a)}{\sqrt{A_{1}(a)+\dot{a}^2}}\right)
\label{e9}
\end{equation}
and
\begin{equation}
p=\frac{- F'(R_0)}{\kappa a}\left( \sqrt{A_{2}(a)+\dot{a}^2}-\sqrt{A_{1}(a)+\dot{a}^2}\right).
\label{e10}
\end{equation}
It is preferable that the shell is made of normal matter, satisfying the weak energy condition, i.e. $\sigma \geq 0$ and $\sigma + p \geq 0$. 
It is important to remark that, in $F(R)$ gravity, $F'(R)>0$ implies that the effective Newton constant $G_{eff} = G/F'(R)$ is positive \cite{bhfr2}, therefore, from a quantum point of view, it prevents the graviton to be a ghost. Further discussion about this topic in wormhole related cases can be found in Ref. \cite{bronnikov}. In what follows, we assume the absence of ghosts, i.e. $F'(R_0)>0$. From Eqs. (\ref{CondGen}), (\ref{e9}), and (\ref{e10}) we can see that the junction conditions require the state equation
\begin{equation}
\sigma -2p=0.
\label{e11} 
\end{equation} 
By combining the time derivative of the equation above with Eqs. (\ref{e9}) and (\ref{e10}), it is easy to verify the conservation equation
\begin{equation} 
\dot{\sigma}+\frac{2\dot{a}}{a}(\sigma + p) = 0,
\label{conservacion1}
\end{equation}
which can be written in the form
\begin{equation}
\frac{d(\sigma \mathcal{A})}{d\tau}+p\frac{d \mathcal{A}}{d\tau}=0,
\label{conservacion2}
\end{equation}
where $\mathcal{A}=4\pi a^2$ is the area of the shell. The first term represents the internal energy change and the second one the work done by the internal forces at the surface $\Sigma$.

\subsection{Stability of static configurations}\label{stab}

For static shells with constant radius $a_0$, from Eq. (\ref{CondGen}) we obtain
\begin{equation} 
-\frac{a_0 A_{1}'(a_0) + 4A_{1}(a_0)}{\sqrt{A_{1}(a_0)}}+\frac{a_0 A_{2}'(a_0)+ 4A_{2}(a_0)}{\sqrt{A_{2}(a_0)}}=0.
\label{CondEstatico}
\end{equation}
The surface energy density $\sigma _0$ and the pressure $p _0$ in this case are given by
\begin{equation} 
\sigma_0=\frac{F'(R_0)}{2\kappa}\left( \frac{A_{2}'(a_0)}{\sqrt{A_{2}(a_0)}}-\frac{A_{1}'(a_0)}{\sqrt{A_{1}(a_0)}}\right)
\label{e13}
\end{equation}
and
\begin{equation}
p_0= \frac{-F'(R_0)}{\kappa a_0}\left( \sqrt{A_{2}(a_0)}-\sqrt{A_{1}(a_0)}\right),
\label{e14}
\end{equation}
respectively.
Now we study the stability of static solutions under spherical perturbations. By taking into account that $\ddot{a}= (1/2)d(\dot{a}^2)/da$ and by defining $z=\sqrt{A_{2}(a)+\dot{a}^2}-\sqrt{A_{1}(a)+\dot{a}^2}$, it is easy to see that Eq. (\ref{CondGen}) can be rewritten in the form
\begin{equation}
az'(a)+2z(a)=0.
\label{CondGen_u}
\end{equation}
By solving this equation, we obtain an expression for $\dot{a}^{2}$ which can be understood in terms of a potential
\begin{equation}
\dot{a}^{2}=-V(a),
\label{condicionPot}
\end{equation}
where 
\begin{equation}
V(a)= -\frac{a_{0}^4\left(\sqrt{A_{2}(a_{0})}-\sqrt{A_{1}(a_{0})}\right)^2}{4a^4} +\frac{A_{1}(a)+A_{2}(a)}{2} -\frac{a^4 \left(A_{2}(a)-A_{1}(a)\right)^{2}}{4 a_{0}^4\left(\sqrt{A_{2}(a_{0})}-\sqrt{A_{1}(a_{0})}\right)^2}.
\label{potencial}
\end{equation}
It is not difficult to verify that $V(a_0)=0$ and, through Eq. (\ref{CondEstatico}), also that $V'(a_0)=0$. The second derivative of the potential at $a_0$ takes the form
\begin{eqnarray}
V''(a_0)&=& -\frac{5 \left(\sqrt{A_{2}(a_{0})}-\sqrt{A_{1}(a_{0}})\right)^{2}}{a_{0}^2}  -\frac{3\left(\sqrt{A_{1}(a_{0})}+\sqrt{A_{2}(a_{0}})\right)^{2}}{a_{0}^2}\nonumber \\
&&-\frac{\left(A_{2}'(a_{0})-A_{1}'(a_{0})\right)^2}{2\left(\sqrt{A_{2}(a_{0})}-\sqrt{A_{1}(a_{0})}\right)^{2}}
-\frac{4\left(\sqrt{A_{1}(a_{0})}+\sqrt{A_{2}(a_{0}})\right)^{2}\left(A_{2}'(a_{0})-A_{1}'(a_{0})\right)}{a_{0}\left(A_2(a_0)-A_1(a_0)\right)}  \nonumber \\
&&+\frac{A_{1}''(a_{0})+A_{2}''(a_{0})}{2}
-\frac{\left(\sqrt{A_{1}(a_{0})}+\sqrt{A_{2}(a_{0})}\right)^{2}\left(A_{2}''(a_{0})-A_{1}''(a_{0})\right)}{2\left(A_{2}(a_{0})-A_{1}(a_{0})\right)}.
\label{potencial2der}
\end{eqnarray}
A static configuration with radius $a_0$ is stable if and only if $V''(a_0)>0$. 
 
\section{Different curvature scalars: the quadratic case}

When we work with different curvature scalars at the sides of the shell, we are restricted to the quadratic case, i.e. $F(R)=R-2\Lambda+\alpha R^{2}$, which does not demand the extra condition of the continuity of $R$ across the surface $\Sigma$, i.e. $[R]\neq 0$ is allowed. In this case, we should only require the continuity of the first fundamental form and of the trace of the second fundamental form, i.e. $[h_{\mu \nu}]=0$ and $[K^{\mu}_{\;\; \mu}]=0$. The derivative $F'(R)=1+2\alpha R$ only depends on the parameter $\alpha$.

\subsection{Construction}

We follow the procedure detailed in the previous section, now with constant $R_1 \neq R_2$. The shell radius $a$ has to satisfy Eq. (\ref{CondGen}). From Eq. (\ref{LanczosQuad}), the field equations become
\begin{equation}
\kappa S_{\mu \nu} =-[K_{\mu\nu}]-2\alpha[RK_{\mu\nu}].
\label{Quad_LanczosGen2}
\end{equation}
so in the orthonormal basis in which $S_{\hat{\imath}\hat{\jmath} }={\rm diag}(\sigma ,p,p)$, we obtain the energy density and the transverse pressure
\begin{equation} 
\sigma= -\frac{2 \ddot{a}+A'_{1}(a)}{2\kappa\sqrt{A_{1}(a)+\dot{a}^2}}\left(1+2\alpha R_{1}\right)+\frac{2 \ddot{a}+A'_{2}(a)}{2\kappa\sqrt{A_{2}(a)+\dot{a}^2}}\left(1+2\alpha R_{2}\right) ,
\label{enden}
\end{equation}
\begin{equation}
p=\frac{\sqrt{A_{1}(a)+\dot{a}^2}}{\kappa a} \left(1+2\alpha R_{1} \right)-\frac{\sqrt{A_{2}(a)+\dot{a}^2}}{\kappa a} \left(1+2\alpha R_{2} \right),
\label{pres}
\end{equation}
respectively. For the reasons explained in the previous section, we assume $F'(R_1)=1+ 2\alpha R_1>0$ and $F'(R_2)=1+ 2\alpha R_2>0$ in order to prevent ghosts. Normal matter at $\Sigma$ should satisfy the weak energy condition. From Eq. (\ref{Tmu}) we can see that $\mathcal{T}_\mu =0$ and thanks to Eq. (\ref{Tg}) the external scalar tension/pressure $\mathcal{T}$ reads  
\begin{equation}
\mathcal{T}=\frac{2\alpha [R]}{\kappa \sqrt{A_{1}(a)+\dot{a}^2}}\left(\ddot{a}+\frac{A'_{1}(a)}{2}+\frac{2}{a}\left(A_{1}(a)+\dot{a}^2\right)\right),
\label{T}
\end{equation}
which by using Eq. (\ref{CondGen}) can be rewritten in the form
\begin{equation} 
\mathcal{T}=-\frac{ 2a\ddot{a}+a A_{1}'+4 \left(A_{1}(a) +\dot{a}^2\right)}{\kappa a\sqrt{A_{1}(a) +\dot{a}^2}}\alpha R_{1}+\frac{2a\ddot{a}+a A_{2}'+4 \left(A_{2}(a) +\dot{a}^2\right)}{\kappa a\sqrt{A_{2}(a) +\dot{a}^2}}\alpha R_{2}.
\label{Trewrit}
\end{equation}
With the help of Eqs. (\ref{enden}), (\ref{pres}), and (\ref{Trewrit}) we find the equation of state that relates $\sigma$, $p$, and $\mathcal{T}$ 
\begin{equation}
\sigma-2p=\mathcal{T}.
\label{state}
\end{equation}
By considering the time derivative of Eq. (\ref{state}) and using Eqs. (\ref{enden}) and (\ref{pres}), we can easily obtain the generalized continuity equation
\begin{equation} 
\dot{\sigma}+\frac{2\dot{a}}{a}(\sigma + p) = \dot{\mathcal{T}},
\label{continuity}
\end{equation}
or equivalently 
\begin{equation} 
\frac{d}{d\tau}(\mathcal{A}\sigma )+ p \frac{d\mathcal{A}}{d\tau} = \mathcal{A} \frac{d\mathcal{T}}{d\tau}.
\label{continuity-bis}
\end{equation}
At the left hand side of this equation, the first term can be interpreted as the change in the total energy of the shell, the second one as the work done by the internal pressure, while the right hand side represents an external flux. The double layer distribution $\mathcal{T}_{\mu \nu }$, obtained from Eq. (\ref{dlay2}), should satisfy
\begin{equation} 
\langle \mathcal{T}_{\mu \nu } ,\Psi ^{\mu \nu } \rangle= - \int_{\Sigma}  \mathcal{P}_{\mu \nu } \left( n^t \nabla _t \Psi ^{\mu \nu} + n^r \nabla _r \Psi ^{\mu \nu } \right),
\label{dlay3}
\end{equation}
for any test tensor field $\Psi ^{\mu \nu }$. The double layer distribution strength, in the orthonormal basis, has components 
\begin{equation}
- \mathcal{P}_{\tau \tau} =\mathcal{P}_{\hat{\theta}\hat{\theta} }=\mathcal{P}_{\hat{\varphi}\hat{\varphi} }=2\alpha[R]/\kappa ,
\label{dlsob}
\end{equation}
which depend on $\alpha $ and $[R]$, so that the dependence of $\mathcal{T}_{\hat{\imath}\hat{\jmath} }$ with the metric is through the unit normal and the covariant derivative.  

\subsection{Stability of static solutions}

For the static solutions, the radius $a_0$ should satisfy Eq. (\ref{CondEstatico}), and the surface energy density $\sigma _0$, the pressure $p_0$, and the external tension/pressure $\mathcal{T}_0$ take the form
\begin{equation} 
\sigma_0=- \frac{A'_{1}(a_{0})}{2\kappa\sqrt{A_{1}(a_{0})}}\left(1+2\alpha R_{1}\right)+\frac{A'_{2}(a_{0})}{2\kappa\sqrt{A_{2}(a_{0})}}\left(1+2\alpha R_{2}\right) ,
\label{enden0}
\end{equation}
\begin{equation}
p_0=\frac{\sqrt{A_{1}(a_{0})}}{\kappa a_{0}} \left(1+2\alpha R_{1} \right)-\frac{\sqrt{A_{2}(a_{0})}}{\kappa a_{0}} \left(1+2\alpha R_{2} \right),
\label{pres0}
\end{equation}
and
\begin{equation}
\mathcal{T}_0=-\frac{ a_0 A_{1}'(a_0) + 4 A_{1}(a_0)}{\kappa a_0 \sqrt{A_{1}(a_0)}}\alpha R_{1} +\frac{a_0 A_{2}'(a_0)+4 A_{2}(a_0)}{\kappa a_0 \sqrt{A_{2}(a_0)}}\alpha R_{2},
\label{T0}
\end{equation}
respectively. In this case, the equation of state results $\sigma_0 -2 p_0 =  \mathcal{T}_0$. The external energy flux vector $\mathcal{T} _{\mu}^{(0)}$ is null and there is a non--zero double layer distribution $\mathcal{T}_{\mu \nu }^{(0)}$ satisfying Eq. (\ref{dlay3}), with $n^t \nabla _t \Psi ^{\mu \nu} =0$ and the strength given by Eq. (\ref{dlsob}). As in the previous section, the stability of the static configurations is determined again by Eq. (\ref{potencial2der}), with $V''(a_0)>0$ corresponding to the stable ones.

\section{Bubbles with charge}\label{charge}

In order to provide an example of the formalism described in the previous sections, we begin with the action
\begin{equation}
S=\frac{1}{2 \kappa}\int d^4x \sqrt{|g|} (F(R)-\mathcal{F}_{\mu\nu}\mathcal{F}^{\mu\nu}),
\label{action} 
\end{equation} 
where $g=\det (g_{\mu \nu})$ and $\mathcal{F}_{\mu \nu }=\partial _{\mu }\mathcal{A}_{\nu } -\partial _{\nu }\mathcal{A}_{\mu }$ is the electromagnetic tensor. In the metric formalism, the field equations obtained from this action, for constant curvature $R$ and considering an electromagnetic potential $\mathcal{A}_{\mu}=(\mathcal{V}(r),0,0,0)$, have the spherically symmetric solution  \cite{bhfr2} in the form given by Eq. (\ref{metric}), where the metric function reads
\begin{equation} 
A (r) = 1-\frac{2M}{r}+\frac{Q^2}{ F'(R) r^2}-\frac{R r^2}{12},
\label{A-metric}
\end{equation}
with $Q$ the charge and $M$ the mass. In this solution, the electromagnetic potential is given by $\mathcal{V}(r)=Q/r$, and the curvature scalar and the cosmological constant are related by $R=4\Lambda$. 
In order to construct a bubble, we adopt $M=0$ and $Q=0$ for the inner region and $M\neq0$ and Q for the outer region.

\subsection{Constant curvature scalar $R_0$}

We begin with a constant curvature scalar $R_0$ at both sides of the shell. Therefore, the metric functions we are going to use have the form
\begin{equation} 
A _1(r) = 1-\frac{R_0 r^2}{12},
\label{metricA1}
\end{equation}
for the inner zone and 
\begin{equation} 
A _2(r) = 1-\frac{2M}{r}+\frac{Q^2}{ F'(R_0) r^2}-\frac{R_0 r^2}{12},
\label{metricA2}
\end{equation}
for the outer one. The possible horizons are determined by the zeros of the  $A_{1,2}(r)$. The geometry given by Eq. (\ref{metricA1}) present a cosmological horizon when $R_0>0$, otherwise it has no horizons. In the case of Eq. (\ref{metricA2}), the metric is singular at $r=0$, and for  $R_0=0$ the horizons are determined by the solutions of a quadratic equation, while for $R_0\neq0$ the horizons are determined by the roots of a fourth degree polynomial.
There exists a critical value of charge $Q_c$ where the number of horizons changes, so it plays an important role. For $R_0>0$ the metric always has the cosmological horizon; in addition, if $0<|Q|<Q_c$ it has the inner and the event horizons, when $|Q|=Q_c$ they fuse into one to finally disappear if $|Q|>Q_c$, a case in which there is a naked singularity at the origin. For $R_0<0$, if $|Q|<Q_c$ the inner and the event horizons are present, when $|Q|=Q_c$ they merge, and if $|Q|>Q_c$ there is a naked singularity and there are no horizons. 

In our construction, the radius $a$ of the bubble is a solution of Eq. (\ref{CondGen}). When $R_0>0$, the value of $a$ should be small enough to avoid the presence of the cosmological horizon coming from the inner metric, but also smaller than the cosmological horizon of the outer part; if $|Q|\le Q_c$ we additionally demand that $a$ should be large enough to remove region inside the event horizon of the outer metric. As discussed above, we take $F'(R_0)>0$; with normal matter preferable at $\Sigma$. In this way, the spacetime, without event or inner horizons, consists of a vacuum region surrounded by a charged thin shell of matter, which in turn is embedded in a region with de Sitter or anti-de Sitter asymptotics, depending on whether $R_0>0$ or $R_0<0$, respectively. 

\begin{figure}[t!]
\centering
\includegraphics[width=0.80\textwidth]{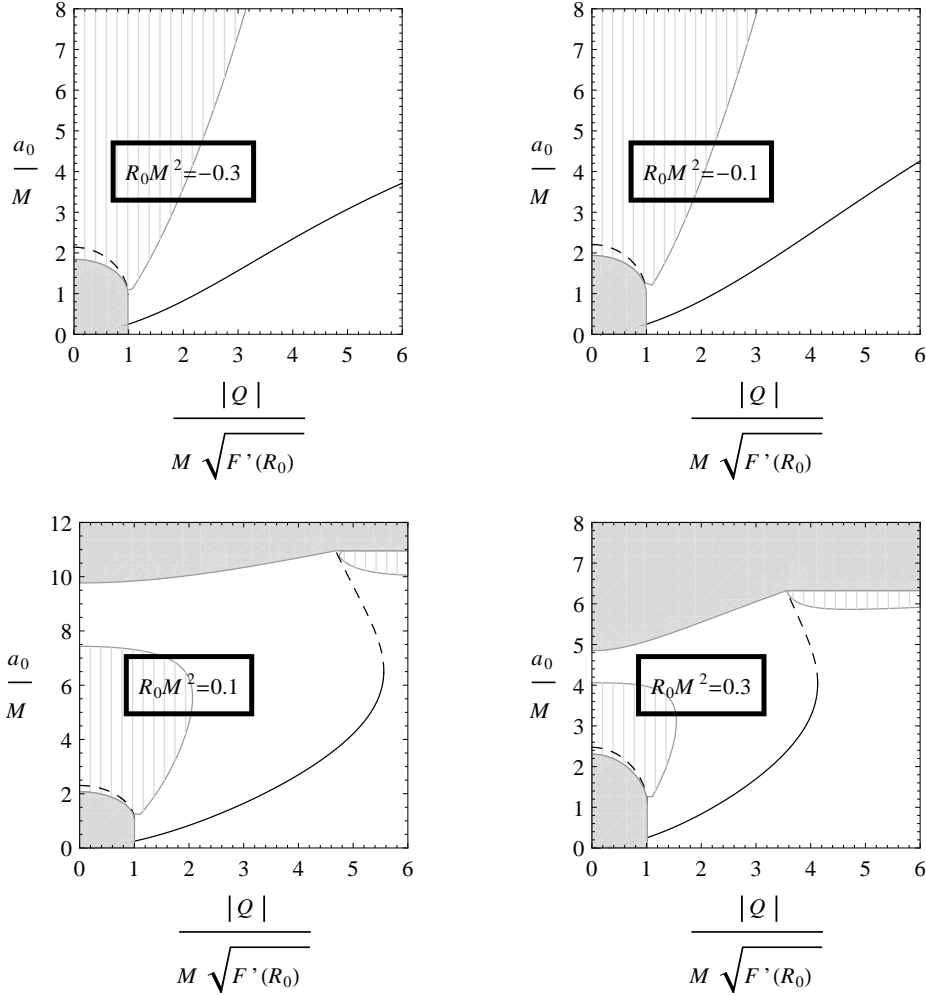}
\caption{Stability of bubbles in $F(R)$ theories for different values of curvature scalar $R_0$. Solid curves represent stable static solutions with radius $a_0$, while dotted curves represent unstable ones. The mass $M$ and the charge $Q$ correspond to the geometry of the outer region. The meshed zones correspond to matter satisfying the weak energy condition and the gray ones are non--physical.}
\label{fig1}
\end{figure}

In particular, for the static configurations, the radius $a_0$ has to be a solution of Eq. (\ref{CondEstatico}) and satisfy the inequalities $\sigma _0 \ge 0$ and $\sigma _0 + p_0 \ge 0$ if the matter is normal. The energy density and the pressure are given by Eqs. (\ref{e13}) and (\ref{e14}), respectively. By using Eq. (\ref{potencial2der}) we can determine the stability of these solutions by recalling that $V''(a_0)>0$ corresponds to stable ones. We present the results graphically in Fig. \ref{fig1}, displaying the most representative of them. All quantities were adimensionalized with the mass, the meshed zones represent shells with normal matter, and the gray ones have no physical meaning. With solid lines we show the stable solutions, while with dotted lines, the unstable ones. The critical charge $Q_c$ is the absolute value of the charge $|Q|$ at the end of the bottom gray zone of the plots. When $R_0<0$ there is an unstable solution for $|Q|$ smaller than $Q_c$, while for larger values of $|Q|$ the solution is stable but requires exotic matter, i.e. it violates the weak energy condition. For  $R_0>0$ there are two unstable solutions, one for small $|Q|$ that extends to the critical charge, while the other one requires exotic matter and is present for large values of $|Q|$; the stable solution appears after the critical charge and exists for a wide range of $|Q|$, but also requires exotic matter. The explicit form of the function $F(R)$, which acts through its derivative $F'(R_0)$, does not affect the qualitative aspects of our results, it only modifies the overall scale, i.e. the quotient $|Q|/\sqrt{F'(R_0)}$ can be interpreted as an effective charge.

\subsection{Different curvature scalars $R_1$ and $R_2$}

By taking into account the procedure of Sec. 3, we construct vacuum bubbles with charge by using metric functions
\begin{equation} 
A _1(r) = 1-\frac{R_1 r^2}{12}
\label{metricA1_Quad}
\end{equation}
and 
\begin{equation} 
A _2(r) = 1-\frac{2M}{r}+\frac{Q^2}{ (1+2\alpha R_2) r^2}-\frac{R_2 r^2}{12},
\label{metricA2_Quad}
\end{equation}
where  $R_1$ and $R_2$ are different constant curvature scalars. The metric defined by Eq. (\ref{metricA1_Quad}) has a cosmological horizon for $R_1>0$, otherwise it has no horizons. In relation to Eq. (\ref{metricA2_Quad}), for $R_2>0$ the cosmological horizon is always present; besides, if $0<|Q|<Q_c$ the metric has the inner and the event horizons, when $|Q|=Q_c$ these horizons coincide, and if $|Q|>Q_c$ both of them vanish resulting in a naked singularity at the origin. For $R_2<0$, if $|Q|<Q_c$ the geometry has the inner and the event horizons which merge when $|Q|=Q_c$, and if $|Q|>Q_c$ there is a naked singularity and there are no horizons. We perform our construction in a similar way to the previous subsection. The radius $a$ of the bubble is a solution of Eq. (\ref{CondGen}) which should be suitably taken to avoid the presence of the inner and the event horizons when $|Q| \le Q_c$. Then the spacetime consists of a vacuum region surrounded by a charged thin shell, which in turn is embedded in a region with de Sitter ($R_2>0$) or anti-de Sitter ($R_2<0$) asymptotics. We adopt $F'(R_1)=1+2\alpha R_1 >0$ and $F'(R_2)=1+2\alpha R_2 >0$; at $\Sigma$ normal matter is desirable.

\begin{figure}[t!]
\centering
\includegraphics[width=0.80\textwidth]{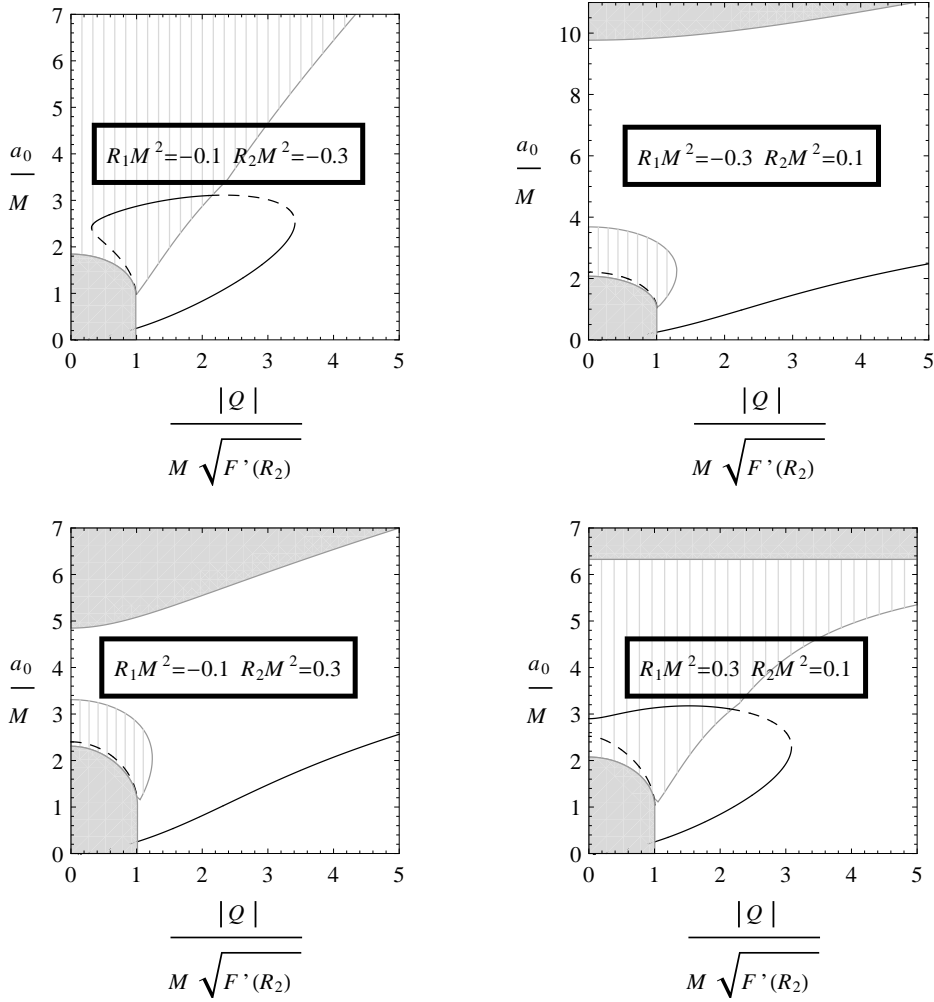}
\caption{Stability of bubbles in quadratic $F(R)$ theories for different values of curvature scalars $R_{1}$ (inner region) and $R_{2}$ (outer region). Solid curves represent stable static solutions with radius $a_0$, while dotted curves represent unstable ones. The mass $M$ and the charge $Q$ correspond to the geometry of the outer region,  $\alpha /M^2=0.1$, and $F'(R_2)=1+2\alpha R_2$. The meshed zones correspond to matter satisfying the weak energy condition and the gray ones are non--physical.}
\label{fig2}
\end{figure}

In the specific case of the static configurations, the radius $a_0$ has to be a solution of Eq. (\ref{CondEstatico}) and fulfill the inequalities $\sigma _0 \ge 0$ and $\sigma _0 + p_0 \ge 0$ for normal matter at the shell. The surface energy density, the pressure, and the external tension/pressure are given by Eqs. (\ref{enden0}), (\ref{pres0}), and  (\ref{T0}), respectively. The shell also presents the dipole layer distribution with a nonzero strength given by Eq. (\ref{dlsob}). We analyze the stability of the solutions by the study of the sign of $V''(a_0)$, and we present some of the results in Fig. \ref{fig2}, in which all quantities are adimensionalized with the mass, the meshed regions represent shells with normal matter, and the gray ones have no physical meaning. The value of the parameter $\alpha$, chosen  as $\alpha /M^2=0.1$ in Fig. \ref{fig2}, does not imprint significant changes in the qualitative behavior of the solutions, it translates into a change of scale. The quotient $|Q|/F'(R_2)$ can be thought as an effective charge. We can observe that the solutions have mainly two different behaviors that depend on the relationship between the value of $R_1$ and $R_2$ instead of the sign of each of them. For $R_1>R_2$ there are stable solutions for a short range of values of charge $|Q|$; these stable solutions exist before and after the critical charge $Q_c$, in the smaller one the shell is composed of normal matter, while the other is exotic. In particular, if $R_1$ and $R_2$ are suitably chosen, there is a stable solution with normal matter for small $|Q|$ and even with no charge, as in the bottom right plot of Fig. \ref{fig2}. For $R_1<R_2$ there are no stable solutions before the critical charge $Q_c$; there is a stable solution with exotic matter for any value of $|Q|$ larger than $Q_c$.

\section{Conclusions}\label{conclu}

We have constructed a large class of spherically symmetric spacetimes with a thin shell of matter, within the framework of $F(R)$ gravity. In arbitrary F(R) theories, spherically symmetric exact solutions are very difficult to obtain without imposing the constant curvature condition. For this reason and also for simplicity, we have adopted geometries with constant curvature scalars at both sides of the shell. In particular, we have considered spacetimes with the same curvature scalar $R_0$, in which there is no restriction for the $F(R)$ function. But in the case with different $R_1$ (inner) and $R_2$ (outer) curvature scalars at the sides of the shell, the junction conditions have limited us to quadratic $F(R)$. For spacetimes with the same $R_0$, the shell has a surface energy density $\sigma$ and an isotropic pressure $p$, related by $\sigma - 2 p = 0$. In the quadratic case with $R_1 \neq R_2$, there is also an external tension/pressure $\mathcal{T}$ related with $\sigma$ and $p$ by $\sigma - 2 p = \mathcal{T}$, a null external energy flux vector $\mathcal{T}_\mu$, and a double layer energy--momentum tensor distribution $\mathcal{T}_{\mu \nu}$ proportional to $\alpha (R_2-R_1)$. This last contribution resembles dipole distributions in classical electrodynamics. In all cases, we have presented a general analysis for the stability of the static configurations under radial perturbations, in terms of a potential.

With the idea to provide concrete examples, we have analyzed spherical bubbles, consisting of a thin shell of matter with mass $M$ and charge $Q$, surrounding vacuum. In these bubbles, we have taken the radius of the shell so that the event horizon (when present) and the region inside it of the outer geometry are removed. In this way, the problems associated to the internal structure of the original geometry, such as the presence of the singularity and the Cauchy horizon (see \cite{pi1990} and the references therein), are avoided in our construction. The spacetime is asymptotically de Sitter or anti-de Sitter, depending on if  the curvature scalar of the outer region is positive or negative, respectively. Unstable static solutions with normal matter at the shell are always present. We have found that there exist stable solutions if the parameters of the model are properly chosen. In the case of constant $R_0$, we have found that the behavior of the solutions basically depends on the sign of $R_0$. For both possible signs there are stable solutions, but these are requiring exotic matter (not satisfying the weak energy condition), for large values of an ``effective'' charge $|Q|/(M\sqrt{F'(R_0)})$  adimensionalized with $M$. In the quadratic $F(R)$ scenario with $R_1 \neq R_2$ the behavior of the solutions depends mainly on the relation between both scalar curvatures, i.e. if $R_1<R_2$ or $R_1>R_2$. When $R_1<R_2$ we have found stable configurations, only with exotic matter, for large values of $|Q|/(M\sqrt{F'(R_2)})$. When $R_1>R_2$ we have obtained stable solutions made of normal matter, for small $|Q|/(M\sqrt{F'(R_2)})$ and also in the absence of charge, for suitable combinations of the parameters.

It is worthy to highlight that  $F(R)$ gravity can be understood as an equivalent of a given scalar--tensor theory; in particular, quadratic $F(R)$ is equivalent to Brans-Dicke theory using a parameter $\omega =0$, where the scalar field $\phi$ and the curvature scalar are related by $\phi=2\alpha R-1$, with a potential $V(\phi)= 2 \Lambda+(\phi ² -2\phi -3)/(4\alpha)$ \cite{dft}. Then our results can be translated to the corresponding scalar--tensor theory.

\section*{Acknowledgments}

This work has been supported by CONICET and Universidad de Buenos Aires.

\end{document}